\documentclass[trackchanges]{aastex7}

\usepackage{subfigure}
\usepackage{float}
\usepackage{graphicx}
\usepackage{multirow}

\usepackage{natbib}
\usepackage{rotating}
\newcommand{\MB}{-19.29}
\newcommand{\MBh}{72.5}
\usepackage{threeparttable}
\begin{document}

\title{The Complete Sample of Available SNe~Ia Luminosity Calibrations from the TRGB Observed with either HST or JWST}

\author[0000-0002-8623-1082]{Siyang Li}
\affiliation{Department of Physics and Astronomy, Johns Hopkins University, Baltimore, MD 21218, USA}

\author[0000-0002-6124-1196]{Adam G.~Riess}
\affiliation{Space Telescope Science Institute, 3700 San Martin Drive, Baltimore, MD 21218, USA}
\affiliation{Department of Physics and Astronomy, Johns Hopkins University, Baltimore, MD 21218, USA}

\author[0000-0002-5259-2314]{Gagandeep S. Anand}
\affiliation{Space Telescope Science Institute, 3700 San Martin Drive, Baltimore, MD 21218, USA}

\author[0000-0002-4934-5849]{Dan Scolnic}
\affiliation{Department of Physics, Duke University, Durham, NC 27708, USA}

\author[0000-0002-8342-3804]{Yukei S. Murakami}
\affiliation{Department of Physics and Astronomy, Johns Hopkins University, Baltimore, MD 21218, USA}

\author[0000-0001-5201-8374]{Dillon Brout}
\affiliation{Departments of Astronomy and Physics, Boston University, Boston, MA 02215, USA}

\author[0000-0001-8596-4746]{Erik R. Peterson}
\affiliation{Department of Physics, Duke University, Durham, NC 27708, USA}

\begin{abstract}

   Distance ladders which calibrate the luminosity of Type Ia supernovae (SNe~Ia) currently provide the strongest constraints on the local value of $H_0$.  Recent studies from the \emph{Hubble Space Telescope (HST)} and \emph{James Webb Space Telescope (JWST)} show good consistency between measurements of SNe~Ia host distances. These are calibrated to NGC 4258 using different primary distance indicators (Cepheids, Tip of the Red Giant Branch (TRGB), J-region Asymptotic Giant Branch, and Miras).  However, some sub-samples of calibrated SNe~Ia employed to measure $H_0$ yield noteworthy differences due to small sample statistics but also due to differences in sample selection.  
  This issue is particularly important for TRGB-derived calibrations owing to the smaller volume they reach compared to Cepheids, reducing sample size and enhancing the size of statistical fluctuations.  To mitigate this issue, we compile the largest and complete (as currently available) sample of \emph{HST} or \emph{JWST} measurements of the TRGB in the hosts of normal SNe~Ia for a total of $N=35$, 50\% larger than the previous largest.  Most are present in the literature, and we compile multiple measures when available.  We also add 5 SNe~Ia hosts from the \emph{HST} archive not previously published.  The full sample together with the Pantheon$+$ SN catalog gives $H_0=72.1-73.3 \pm 1.8$ km/s/Mpc (depending on methodology), in good agreement with the value of 72.5 $\pm 1.5$ km/s/Mpc from \emph{HST} Cepheids in hosts of 42 SNe~Ia calibrated by the same anchor, NGC 4258.  We trace the difference in the result of $H_0=70.4 \pm 1.9$ km/s/Mpc from \citealt{Freedman_JWST_H0_2024arXiv240806153F} to 11 hosts not selected for that CCHP compilation (of $N=24$) which alone yield $H_0=74.1$ km/s/Mpc, 2$\sigma$ higher than the selected sample.  A smaller increase of 0.6 km/s/Mpc comes from a commonly employed correction for peculiar velocities.  
  
\end{abstract}



\section{Introduction}

The most precise route to the Hubble constant ($H_0$) uses primary distance indicators such as Cepheids, Tip of the Red Giant Branch (TRGB), J-region Asympototic Giant Branch (JAGB), and Miras to calibrate the fiducial luminosity of standardized Type Ia supernovae (SNe~Ia).  With good agreement demonstrated between the current generation of primary distance indicators (\citealt{Riess_2022ApJ...934L...7R, Riess2024ApJ...977..120R, Freedman_JWST_H0_2024arXiv240806153F}, here R22 and F25, respectively), 
variations in $H_0$ are dominated by the small sample size of calibrated SNe~Ia.   With each SN~Ia having an intrinsic scatter of $\sim$7\% in $H_0$, one needs a sample of $>$ 25 to reduce 1$-$2 $\sigma$ fluctuations to 1$-$2 km/sec/Mpc.  

SN samples have been calibrated by Cepheids largely due to the greater volume reached with 42 SNe~Ia \citep{Riess_2022ApJ...934L...7R}. That sample is complete in distance (D$\sim$40 Mpc or $z \sim 0.01$) to the year 2021.   For TRGB, samples have surpassed $\mathcal{O}$(10) more recently. \cite{Jang_2015ApJ...807..133J, Jang_2017ApJ...836...74J} (JL17) compiled eight SNe~Ia to find $H_0=71.7 \pm 2.6$ km/s/Mpc, or alternatively 73.7 $\pm$ 2.8 km/s/Mpc from six SN~Ia with low-reddening (those sufficient to pass common SNe~Ia quality cuts).  This sample grew in \cite{Freedman_2019ApJ...882...34F} (F19) to 18, resulting in $H_0$ = 69.8 $\pm$ 1.9 km/s/Mpc, or as reanalyzed by the Extragalactic Distance Database team (EDD; \citealt{Anand_EDD_2021AJ....162...80A, Anand_2022ApJ...932...15A}, A22), 71.5 $\pm$ 1.8 km/s/Mpc calibrated directly to the masers in NGC 4258 \citep{Reid_2019ApJ...886L..27R}. The Comparative Analysis of TRGBs (CATs) team remeasured these (adding additional galaxies from the archive) using a contrast ratio approach to standardizing the TRGB magnitudes and an unsupervised tip detection algorithm and found  $H_0$ = 73.2 $\pm$ 2.1 km/s/Mpc \citep{Scolnic_2023ApJ...954L..31S}.  The availability of the \emph{James Webb Space Telescope (JWST)} provided a new platform for TRGB measurements. \cite{Li_TRGB_vs_Ceph_2024arXiv240800065L} (L24) calibrated 10 SN~Ia with \emph{JWST}, with some overlap of the prior \emph{Hubble Space Telescope  (HST)} TRGB sample, which gave $H_0$ = 74 km/s/Mpc, and F25 calibrated 11 with \emph{JWST}, also with some prior \emph{HST} overlap, that gave $H_0$ = 69 km/s/Mpc.   Importantly, these two \emph{JWST} samples have little overlap and the difference in $H_0$ between these two samples is largely matched in HST Cepheid observations of the same two subsamples \citep{Riess_JWST_H0_2024arXiv240811770R}.  A revision of F25 produced an expanded sample of TRGB measures from either \emph{HST} or \emph{JWST} in the hosts of $N=24$ SNe~Ia resulting in $H_0$ = 70.4 $\pm$ 1.9 km/s/Mpc. However, that study does not include all \emph{HST} or \emph{JWST} data available in the archive or literature nor provides selection criteria that would explain exclusions or otherwise comparable data.
The goal of this work is to collect the complete\footnote{Complete defined as available rather than to a limiting distance, a consequence of the selections of disparate observing programs.} sample of all presently available \emph{HST} or \emph{JWST} measures to increase the sample, study internal agreement, and reduce sample size fluctuations.

We define a maximal uniform sample comprising all hosts of spectroscopically normal SNe~Ia with observations suitable for TRGB measurements, either published or presently available (in the Spring of 2025) in the archive, using \emph{HST} or \emph{JWST}, consistently calibrated to NGC 4258. While SN and TRGB quality varies, we initially include all examples accepted in primary literature and introduce quality as a study criterion later. Several host galaxies also have unpublished \emph{HST} archival observations from past years which we can measure and include. This sample currently includes $N=35$, $\sim$50\% larger than any previously used for $H_0$ determination. 

In Section \ref{sec:TRGB_from_the_archive}, we measure the TRGB in five hosts of spectroscopically normal SN~Ia that have data publicly available in the \emph{HST} but not included in a previous TRGB \emph{HST} $H_0$ sample.  In Section \ref{sec:TRGB_compilation}, we compile a table of TRGB distances from the literature corresponding to all 35 SNe~Ia available to measure $H_0$ with. We also show how $H_0$ varies with different SNe~Ia subsample selection and discuss their implications.

\section{TRGB from the Archive} \label{sec:TRGB_from_the_archive}

We identify five galaxies with haloes that have been observed by \emph{HST} Advanced Camera for Surveys (ACS) and have both \emph{F606W} and \emph{F814W} images publicly available on the Mikulski Archive for Space Telescopes (MAST)\footnote{\url{https://mast.stsci.edu/search/ui/\#/}}. In addition, these galaxies have hosted SNe~Ia but do not yet have a corresponding published TRGB measurement in the literature: NGC 3982, NGC 4414, NGC 4639, and NGC 4666 from GO-17079 \citep[PI: I. Jang; ][]{Jang_2022hst..prop17079J} and NGC 4457 from GO-16453 \citep[PI: K. McQuinn;][]{McQuinn_2020hst..prop16453M}. We use these to augment the host galaxy sample used to measure a TRGB-based $H_0$. 

\begin{figure*}[htbp]
  \centering
  \begin{tabular}{cc}
    \includegraphics[width=0.45\linewidth]{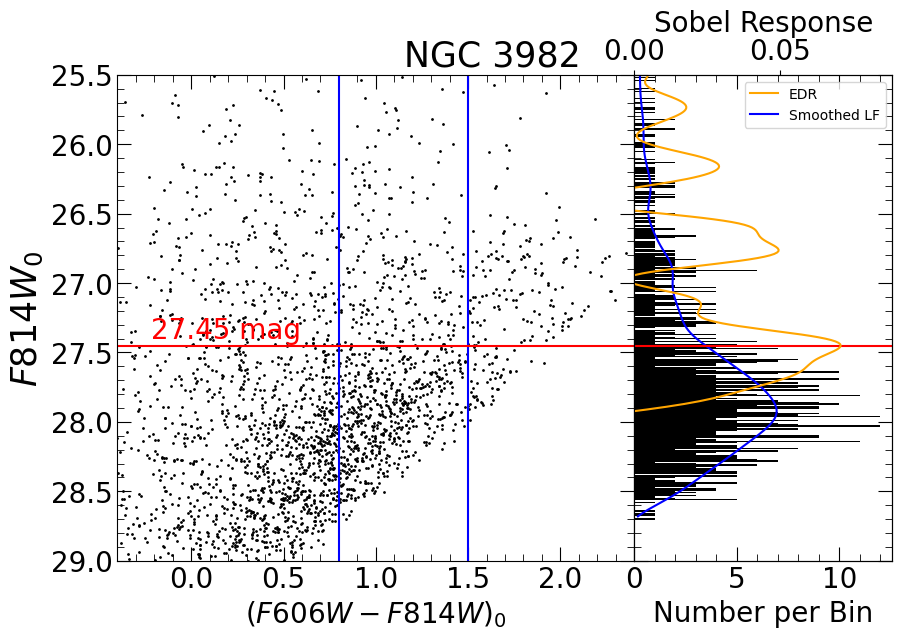} &
    \includegraphics[width=0.45\linewidth]{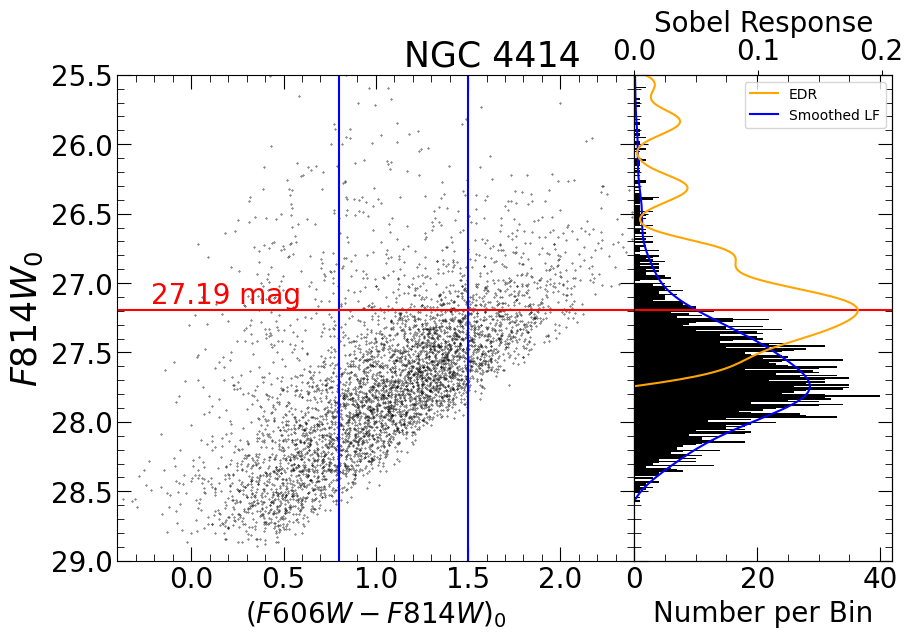} \\
    \includegraphics[width=0.45\linewidth]{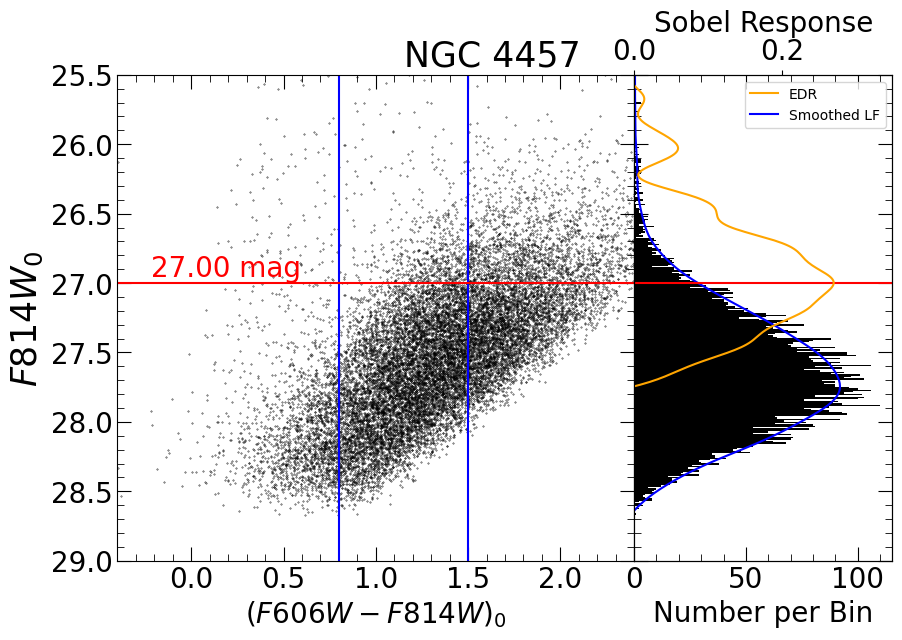} &
    \includegraphics[width=0.45\linewidth]{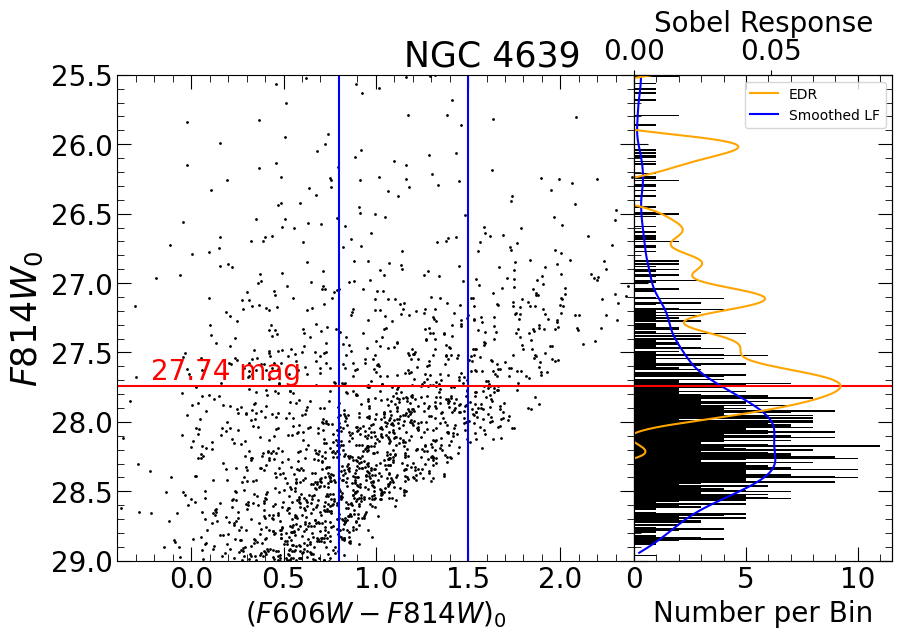} \\
    \multicolumn{2}{c}{\includegraphics[width=0.45\linewidth]{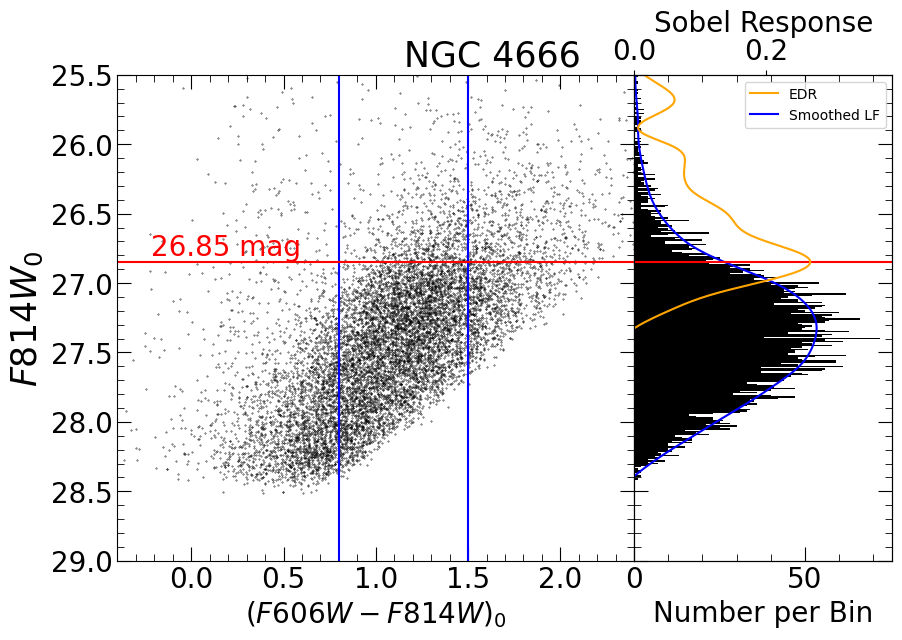}} \\  
  \end{tabular}
 {\footnotesize  \caption{Color magnitude diagrams and luminosity functions for the five host galaxies analyzed in this study. The left panels of each plot show the \emph{F814W} vs. \emph{F606W - F814W} color-magnitude diagrams for each galaxy, with blue lines corresponding to the color cuts used for the TRGB measurements. Magnitudes shown are values after applying foreground extinction corrections. The right panels show the luminosity functions of stars after applying color cuts, with the smoothed luminosity functions shown with the blue lines. The Sobel edge-detector responses are shown by the orange lines. We annotate the measured TRGB magnitudes in red.}}
  \label{fig:CMDs}
\end{figure*}

We begin by retrieving the publicly available \texttt{*.flc} \emph{F606W} and \emph{F814W} images from these programs from MAST. Because the observations from these programs were taken across several orbits, the MAST pipeline produces separate drizzled images corresponding to each visit, resulting in multiple drizzled images per epoch that divides the exposure times across these images. Before performing photometry on the \texttt{*.flc} images, we aim to create the deepest reference image possible. We first align the WCS for all the \emph{F606W} and \emph{F814W} \texttt{*.flc} images for each galaxy using \texttt{tweakreg}, then drizzle the \emph{F814W} images together using \texttt{Astrodrizzle} \citep{2015ASPC..495..281A}. We perform photometry on the \texttt{*.flc} images using the DOLPHOT software 
\citep{Dolphin_2000PASP..112.1383D, Dolphin_2016ascl.soft08013D}, with the newly drizzled images as reference frames. We use the DOLPHOT parameters provided in \cite{Williams_2014ApJS..215....9W} and apply DOLPHOT quality cuts based on the works of \cite{McQuinn_2017AJ....154...51M} and A22: $(Crowd_{F606W} + Crowd_{F814W}) < 0.8$, $(Sharp_{F606W} + Sharp_{F814W}) ^ 2 \leq 0.075$, $Type \leq 2$, $SNR_{F606W, F814W} \ge 3$, and $Flag_{F606W, F814W} = 0$. We show the CMDs for these observations in Fig. \ref{fig:CMDs}. We apply foreground extinction corrections using \emph{E(B-V)} values from \cite{Schlafly_2011ApJ...737..103S} and adopt the \cite{Fitzpatrick_1999PASP..111...63F} $R_v$ = 3.1 reddening law with $A_{\lambda}/A_V$ = 1.725 and 2.799 for \emph{F814W} and \emph{F606W}, respectively, remaining consistent with A22. 

\begin{deluxetable*}{ccccccccccccc}[htbp]
\tabletypesize{\scriptsize}
\caption{$D_{25}$ Ellipse Parameters and TRGB Measurements}
\scriptsize
\label{tab:spatial_cuts}
\tablehead{
\colhead{Galaxy} & \colhead{Program} & \colhead{Observation Date} & 
\colhead{RA} & 
\colhead{Dec} & 
\colhead{PA [\textdegree]} & 
\colhead{Axis Ratio} & 
\colhead{SMA [arcsec]} & \colhead{$m_{TRGB}$} & \colhead{$\sigma$} 
}
\startdata
NGC 3982 & 17079 & 2023-10-03, 2023-10-05  & 11h56m28.1280s & +55d07m30.766s & 38 & 0.898 & 52.83  & 27.45 & 0.13 \\
NGC 4414 &17079 & 2024-04-19, 2024-04-09, 2024-04-11& 12h26m27.1491s & +31d13m24.694s & 155 & 0.562 & 108.9  & 27.19& 0.08 \\
NGC 4457 &16453 & 2021-04-14, 2021-04-17, 2021-04-18, 2021-07-27& 12h28m59.0203s & +03d34m14.062s & 75 & 0.851 & 80.75 & 27.00 & 0.09\\
NGC 4639 &17079 & 2023-06-23, 2023-06-24, 2023-06-25, 2024-05-22& 12h42m52.3879s & +13d15m26.784s & 123 & 0.676 & 82.65 & 27.74 & 0.08\\
NGC 4666& 17079 &2023-07-10, 2023-07-01  & 12h45m08.6345s & -00d27m43.290s & 42 & 0.282 & 137.15 & 26.85 & 0.02\\
\enddata
\tablecomments{Summary table for elliptical cuts used to define the halo and TRGBs measured in this study. We measure the TRGB using stars that lie outside these ellipses.}
\end{deluxetable*}

The TRGB marks the onset of the helium flash in red giant stars \citep{Iben_1983ARA&A..21..271I}, and the magnitude at which this occurs is visible as a discontinuity in the giant branch luminosity function. The TRGB can be measured using several ways; for instance, by fitting a broken power law model \citep{Mendez_2002AJ....124..213M} to the luminosity function using maximum likelihood estimation \citep{Makarov_2006AJ....132.2729M, Li_2022ApJ...939...96L, Li_2023ApJ...950...83L} or least-squares fitting \citep{Wu_2014AJ....148....7W, Cronjevic_2019}. Another approach is to run an edge-detector (e.g. a Sobel filter; \citealt{Lee_1993ApJ...417..553L}) across the luminosity function to trace its first derivative and identify the location of maximum change \citep[see, for instance, ][]{Hatt_2017ApJ...845..146H, Wu_2023ApJ...954...87W}.

Our goal is not to explore methodological differences in the measure of the TRGB but rather to improve sample statistics by compiling a complete sample of all available hosts.  To that end, 
 we maintain consistency with the measurement method from F19 and H21 for the five hosts with archival data by adopting similar procedures.  Therefore,  we use a Sobel-filter based approach to measure the TRGB, similar to that described in \cite{Hatt_2017ApJ...845..146H} and use the same calibration of $M_{TRGB}$ = $-$4.049 $\pm$ 0.015 (stat) $\pm$ 0.035 (sys)~mag from \cite{Jang_2021ApJ...906..125J} for NGC 4258.   We use a fixed color range of 0.8~mag $<$ \emph{F606W} $-$ \emph{F814W} $<$ 1.5~mag, a color range where the TRGB can be approximated to be flat with color following \cite{Jang_2017ApJ...835...28J}. We apply spatial cuts using the 25th magnitude B-band isophotal radius ($D_{25}$), from the parameters available from the NASA Extragalactic Database (NED)\footnote{\url{https://ned.ipac.caltech.edu/}} and listed in Table \ref{tab:spatial_cuts} and exclude stars that fall inside the ellipse. We adopt a smoothing scale for the Gaussian-weighted Locally Estimated Scatterplot Smoothing \citep[GLOESS;][]{Hatt_2017ApJ...845..146H} of 0.1~mag.  We also apply Poisson weighting to the Sobel filter output to be consistent with F19, F25, and \cite{ Freedman_Tensions_2021ApJ...919...16F}, but acknowledge criticism that this can bias measurements in some cases, see \citealt{Anderson_2024ApJ...963L..43A, Anand_2024ApJ...966...89A}. We estimate errors on these TRGB measurements using 10,000 bootstrap resamples. We list the measured TRGB magnitudes and their errors in Table \ref{tab:spatial_cuts}.

\section{TRGB compilation} \label{sec:TRGB_compilation}

\subsection{Complete TRGB Sample}

In Table \ref{tab:distances}, we compile the TRGB distance measures for the complete sample of hosts of 35 SNe~Ia obtained with {\it HST} and {\it JWST}, consistently calibrated to NGC 4258 as provided by the indicated literature sources.  
Where available, we also provide a secondary source of TRGB host measurements  (see, for instance, \citealt{Jang_2017ApJ...835...28J, Yuan_2019ApJ...886...61Y, Anand_EDD_2021AJ....162...80A, Anand_2021MNRAS.501.3621A, Anand_2022ApJ...932...15A, Li_2022ApJ...939...96L, Li_2023ApJ...950...83L, Li_CATs_2023ApJ...956...32L, Freedman_JWST_H0_2024arXiv240806153F}). Some of these studies use a different measurement method; for instance, A22 measures the TRGB in NGC 4258 and in SNe~Ia hosts by fitting a model luminosity function instead of using edge-detection.  The distances taken from the literature in this table sometimes do not include any NGC 4258 error (which would include both the NGC 4258 measurement error and maser distance error), only the NGC 4258 TRGB measurement error, or the full NGC 4258 errors; these are listed in the table notes. When comparing two distances derived using \emph{HST} and \emph{JWST}, it is important to ensure that the TRGB measurement error in NGC 4258 is included but the maser distance error is not, as the maser distance error is common to both distances. The same applies to comparing distances for the same instrument (i.e. two \emph{HST} distances or two \emph{JWST} distances); in this case, the full NGC 4258 errors (maser distance and tip measurement uncertainty) should be removed as they are shared. We take this into account for our $H_0$ variants, described later, and in Fig. \ref{fig:TRGB_diffs}. We explore the impact of some of these other measures on $H_0$ in the next section.

The standardized SN magnitudes in Table \ref{tab:distances} come from the Pantheon$+$ compilation \citep{Scolnic_Pantheon_2022ApJ...938..113S}, with mean magnitudes taken from Table 6 of R22 or Table 2 from \cite{Scolnic_2023ApJ...954L..31S} for SNe not in R22\footnote{A small discrepancy between SN magnitudes listed in R22 Table 6 and 
\cite{Scolnic_2023ApJ...954L..31S} exists only for SNe in Pantheon$+$ with 2 data sources and is caused by the use in R22 Table 6 of their mean and in S23 by their (IDL) median, the larger of the two, amounting to a mean difference in S23 of 0.24 km/s/Mpc.}.  For three SNe Ia used here and not in these studies, SN 2021J, SN 2020nvb, and SN ASASSN-14lp, we produce the standardized magnitudes using the available light curves from YSE, SWIFT, and CSP (Shappee et al. 2016), respectively.

\subsection{Internal Comparisons}

\begin{figure}[htbp]
    \centering
    \includegraphics[width=1\linewidth]{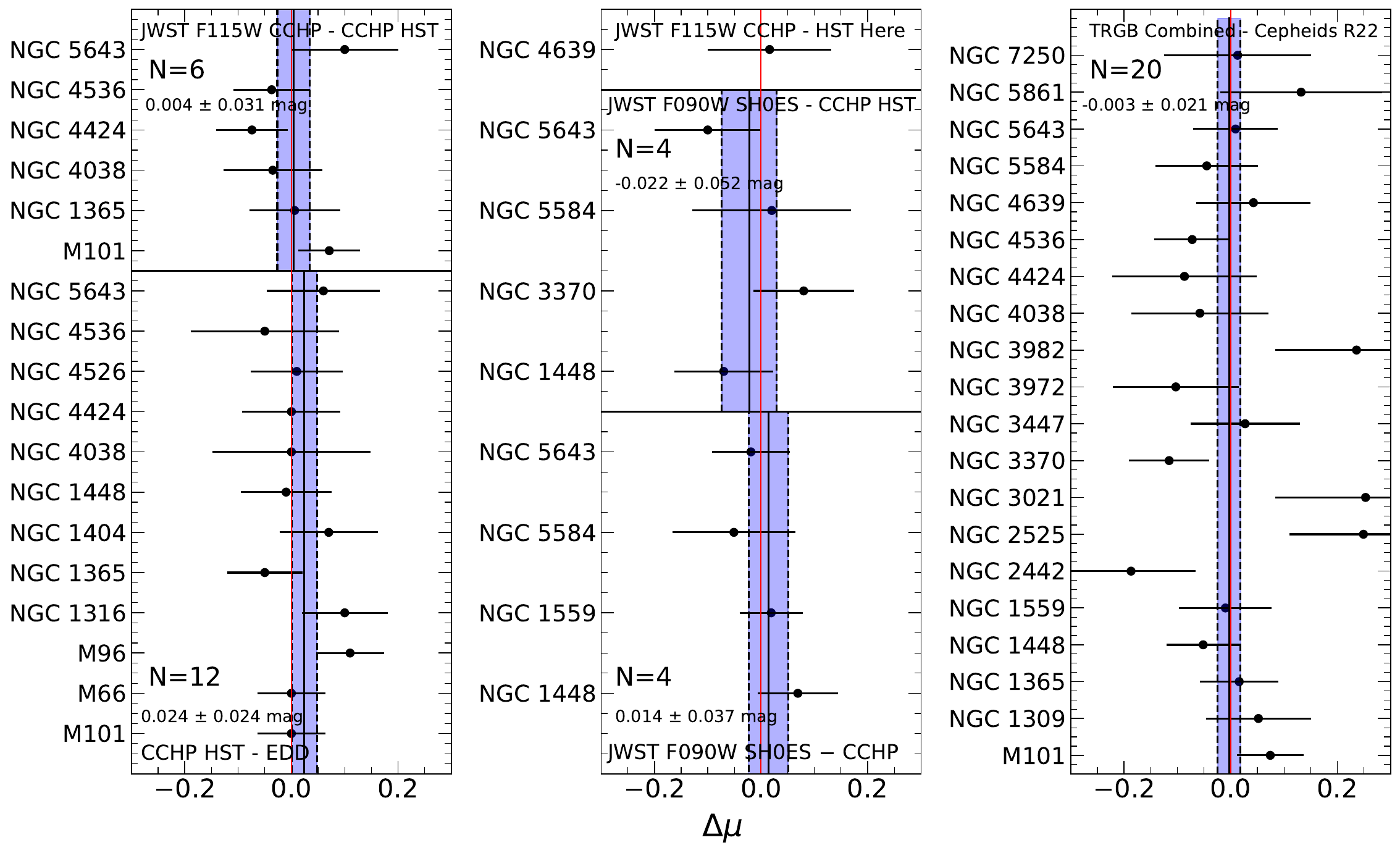}
    \caption{\footnotesize Differences between TRGB distances to host galaxies from various sources listed in Table \ref{tab:distances}. We plot the weighted means and error on the weighted means for each group with solid and dashed lines, respectively. We also place a red line at $\Delta\mu=$0.00~mag for reference. We include labels for which datasets were used: {\it JWST} F115W CCHP refers to the distances from \cite{Freedman_JWST_H0_2024arXiv240806153F}, CCHP HST from F19 and F25, EDD from A22, HST here from Table \ref{tab:distances}, JWST F090W SH0ES from L24, CCHP (bottom-most section of the center subplot) from \cite{Freedman_JWST_H0_2024arXiv240806153F}, TRGB combined from all galaxies that have both Cepheid and TRGB distances (where we combined HST and JWST measures), and Cepheids from R22. For these comparisons, we remove the NGC 4258 TRGB measurement error from the JWST \emph{F090W} SH0ES distances in the JWST \emph{F090W} SH0ES 
 - CCHP comparison to remain consistent with CCHP. We also remove the NGC 4258 error (maser and TRGB measurement errors) from the distances for NGC 1380 and NGC 7814 to remain consistent with the other distances.}
    \label{fig:TRGB_diffs}
\end{figure}


We compare TRGB measurements for the same hosts between studies and telescopes, listed in Table \ref{tab:distances}, in Fig. \ref{fig:TRGB_diffs}. These differences include comparing the CCHP \emph{JWST} and \emph{HST} distances, CCHP and EDD distances using the same data from \emph{HST}, \emph{JWST} CCHP and the \emph{HST} galaxies analyzed here (single target), \emph{JWST} SH0ES measurements and the CCHP \emph{HST} galaxies, \emph{JWST} SH0ES and CCHP distances, and \emph{HST} Cepheids and combined \emph{HST} and \emph{JWST} TRGB distances. None of these comparisons yield a significant difference, with results given in Figure \ref{fig:TRGB_diffs}. 

Given the good agreement between \emph{HST} and \emph{JWST}-based distance measures (both calibrated to NGC 4258), 
it is reasonable to combine TRGB sample measures to produce a complete \emph{HST}+\emph{JWST} TRGB sample.   To produce this, we take the weighted mean of the two distances, \emph{HST} and \emph{JWST}, when both measures are available and use that as a baseline distance. We take care to exclude the maser distance or a common measure of the TRGB in NGC 4258 from the averaging before reintroducing the common NGC 4258 tip error. For the first baseline we adopt the \emph{HST} measures on the F19 system and the \emph{JWST} measures provided by F25 or L24 (with the latter consistent with these measured reproduced for SH0ES galaxies by \citealt{Hoyt_2025arXiv250311769H} (H25) at the 0.01~mag level).  We also provide variants to these measures in Table \ref{tab:distances}, including the the reanalysis of F19 from A22 and the CCHP reanalysis of \emph{JWST} SH0ES observations from F25.

 We also compare the differences between these mean compiled TRGB distances to host galaxies with the \emph{HST} Cepheid distances from R22 anchored to NGC 4258 only (also listed in \citealt{Riess2024ApJ...977..120R}), and show this in the right most subplot of Fig. \ref{fig:TRGB_diffs}. The result is a weighted mean of $\Delta\mu(TRGB-Cepheids)$ = $-$0.003 $\pm$ 0.021 (stat)~mag for the \emph{HST} TRGB sample (including those measured here), and hence no statistically significant difference.  This is the largest comparison to date between TRGB and HST Cepheid measurements for SN hosts with $N=20$ objects.  We note a similar result from F25 comparing $N=14$ of 0.025 $\pm$ 0.021 mag, however anchoring the \emph{HST} Cepheids to (only) NGC 4258, for a more direct comparison to TRGB anchor the same way, would reduce even that difference to 0.010 $\pm 0.021$ mag.

\begin{deluxetable*}{|l|l|c|c|c|}  
\tablecaption{$H_0$ Measurement Variants}
\tabletypesize{\normalsize}
\tablehead{
\colhead{Variant} &\colhead{$N_{SNe}$}  &\colhead{$M_B$}  & \colhead{$H_0$}  & \colhead{$\sigma$} 
}
\startdata
All (baseline) HST or JWST Galaxies Combined & 35 & $-$19.303 & 72.1 & 1.1\\
EDD instead of F19 or H21 Measures for HST & 35 & $-$19.266 & 73.3 & 1.3\\
Only CCHP JWST & 11 & $-$19.362 & 70.2 & 1.8\\
Only SH0ES JWST & 10 & $-$19.257 & 73.6 & 1.8\\
All JWST (CCHP and SH0ES) & 19 & $-$19.292 & 72.4 & 1.6\\
``CCHP Selected" (F19/J17/F24, no 2021pit) & 24 & $-$19.344 & 70.7 & 1.2 \\
``CCHP Selected" (no vel. corr.) & 24 & $-$19.337 & 70.2 & 1.2 \\
CCHP Not Included & 11 & $-$19.242 & 74.1 & 1.7\\
Baseline w/o J17 (removes irreproducible edge detections) & 33 & $-$19.312 & 71.8 & 1.1\\
Remove SN which fail QC & 30 & $-$19.298 & 72.2 & 1.1\\
All (baseline) w/o biggest pull host, NGC 1316 & 32 & $-$19.288 & 72.6 & 1.1\\
\enddata
\tablecomments{$H_0$ values calculated using several combinations of SN~Ia magnitudes and host galaxy distances, to compare with the baseline result using N=35 SNe~Ia. The errors in $H_0$ in this table, provided in the right most column, are calculated only using the weighted mean error on $M_B$; the full error in $H_0$ will include 0.032~mag from the maser distance from \cite{Reid_2019ApJ...886L..27R} and 0.006~mag from the error in 5$a_b$ \citep{Riess_2022ApJ...934L...7R}. For the `Only CCHP JWST' variant, we include the \emph{F115W} TRGB distance to NGC 5643 from \cite{Freedman_JWST_H0_2024arXiv240806153F} of 30.643 $\pm$ 0.071~mag. For the ``All JWST (CCHP and SH0ES) variant, we use the SH0ES distance to NGC 5643 rather than the CCHP distance due to preference with using the \emph{F090W} TRGB over the \emph{F115W} TRGB due to a weaker color dependence. For all variants, we include the NGC 4258 TRGB measurement errors in host galaxy distances.}
\label{tab:variants}
\end{deluxetable*}

\begin{figure}[htbp]
    \centering
    \includegraphics[width=1\linewidth]{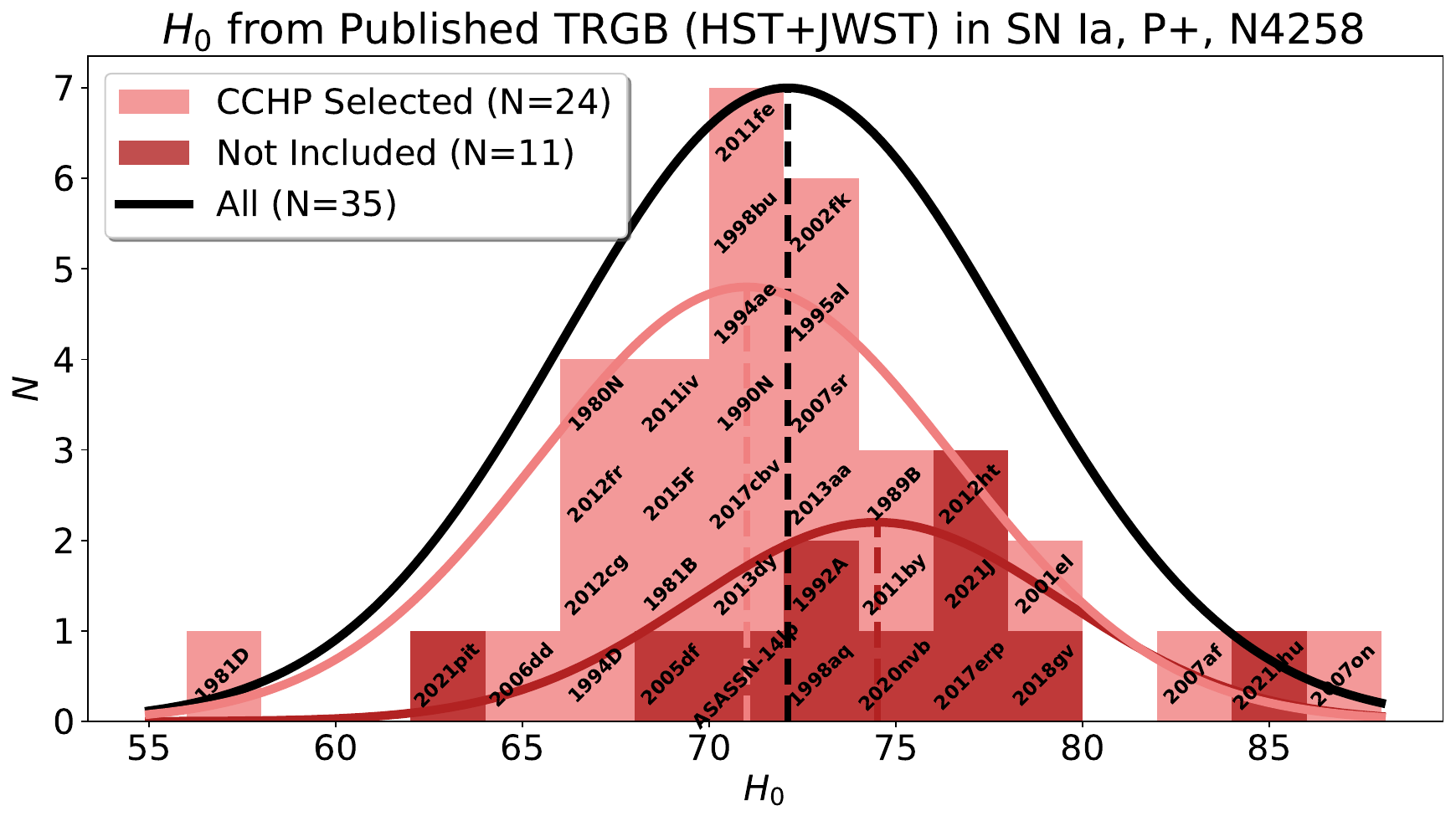}
    \caption{\footnotesize Histogram of $H_0$ from individual SN~Ia, showing 35 SN~Ia and the subsample of 11 SN~Ia left out in \cite{Freedman_JWST_H0_2024arXiv240806153F} (24 included in \cite{Freedman_JWST_H0_2024arXiv240806153F} shown in pink). Three Gaussians correspond to three different $H_0$ values as listed listed in Table \ref{tab:variants}: ``CCHP Selected (F19/J17/F24, no 2021pit)" (left; N=24), ``All (baseline) HST or JWST Galaxies Combined" (center; N=35), and ``CCHP Not Included" (right; N=11). The dashed lines correspond to the means of the Gaussians.}
    \label{fig:H0_histogram}
\end{figure}


\begin{figure}[htbp]
    \centering
    \includegraphics[width=1\linewidth]{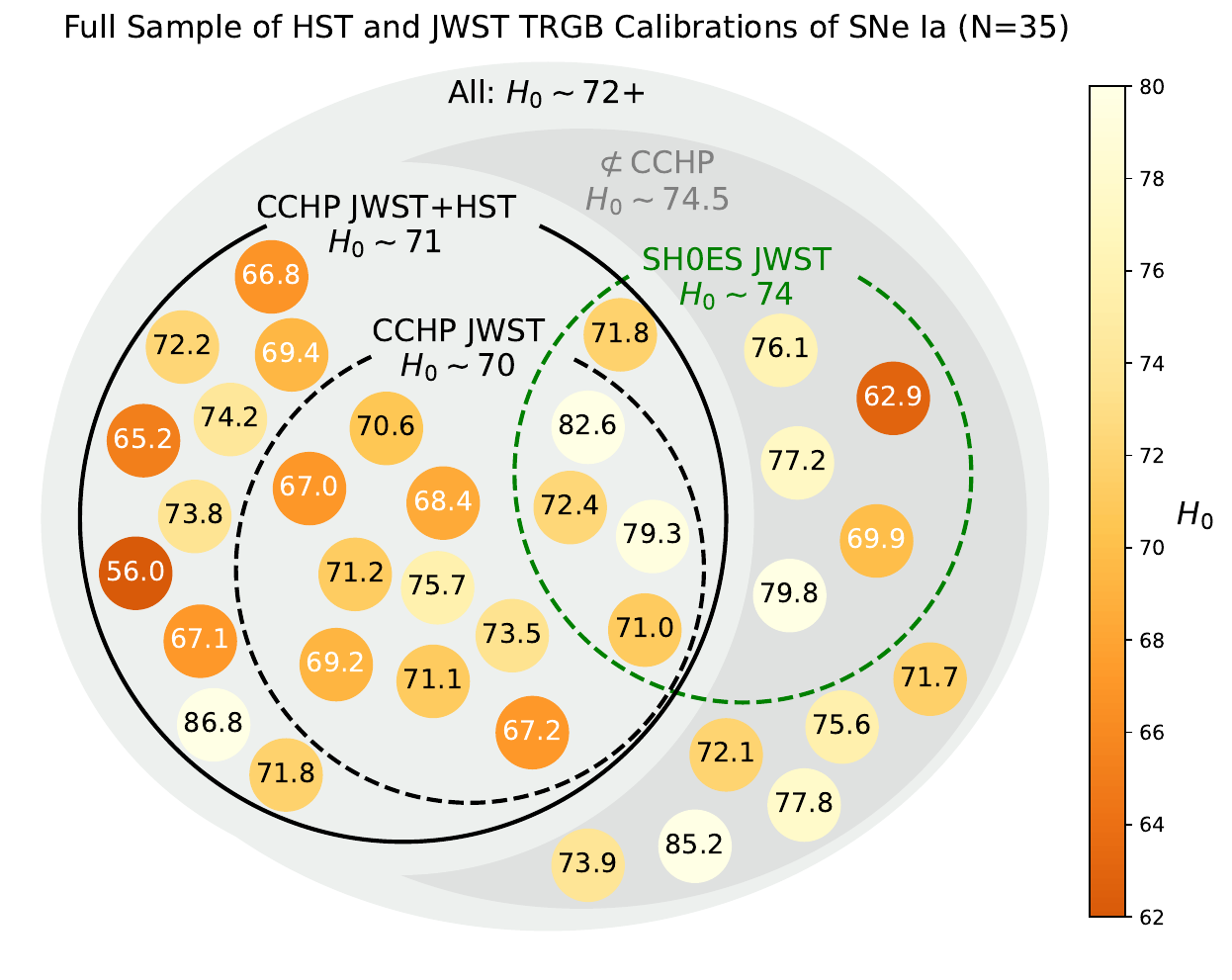}
    \caption{\footnotesize A Venn diagram of SNe Ia contained in the TRGB hosts covered by different sub-samples and the $H_0$ values inferred by each TRGB-calibrated SNe~Ia. The shade in color, from dark brown to light yellow, corresponds to lower to higher values of $H_0$.  Due to the individual SN~Ia scatter, differences in $H_0$ are produced by different subsamples.}
    \label{fig:marble_plot}
\end{figure}


\begin{figure}[htbp]
    \centering
    \includegraphics[width=1\linewidth]{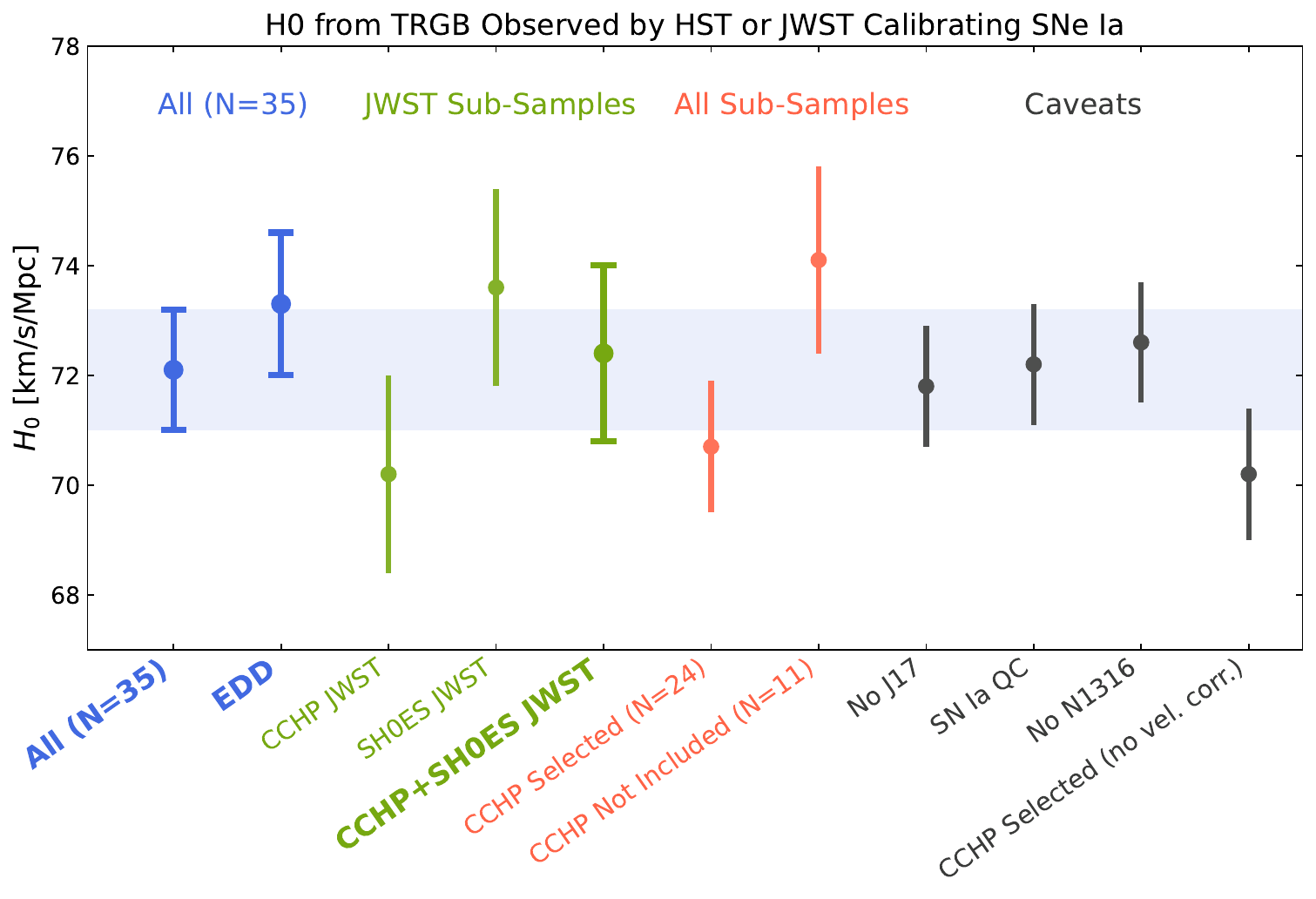}
    \caption{\footnotesize The $H_0$ variants listed in Table \ref{tab:variants}. All of theses values use the same underlying distances to host galaxies as listed in Table \ref{tab:distances}. Variants using the full set of 35 SN~Ia fall around $H_0$ $\sim$72 km/s/Mpc; smaller subsamples, indicated by thinner lines without caps, result in fluctuations around this value. The common maser distance error of 1.5\% is not included here.}
    \label{fig:H0_whisker}
\end{figure}


\subsection{Sub-sample $H_0$} \label{sec:subsamples}

The full sample of 35 SNe~Ia currently provides the highest statistical leverage in measuring $H_0$ and for investigating the variations in presentations of past TRGB $H_0$ measurements. In Table \ref{tab:distances}, we provide the name of the SN~Ia in each host and the standardized magnitude, $m_b^0$ from the Pantheon $+$ SN compilation following \cite{Scolnic_Pantheon_2022ApJ...938..113S}, using the {\it mean SN mag} if multiple surveys are available for a given SN.  $H_0$ follows from
\begin{equation}
5\, \textrm{log}({\rm H}_0/\MBh)=M_B^0-(\MB)\, 
\end{equation} based on the distance ladder fits from F22 which use NGC 4258 as the sole anchor and a Hubble flow sample of all host types (in recognition that the TRGB hosts include late and early type hosts).  We note that excluding peculiar velocity corrections, fit 57 in R22 (derived from 2M++, see \citealt{Peterson_2022ApJ...938..112P}) shifts the Hubble diagram intercept for Pantheon $+$ by 0.015 mag, equivalent to a substitution of the reference value of $M_B$ in equation 1 to $-$19.275 mag.

In Tables \ref{tab:distances} and  \ref{tab:variants}, we show values of $H_0$ calculated using individual SN~Ia and different combinations of SNe~Ia, respectively. Using all 35 SNe~Ia, calibrated using the weighted mean of the baseline \emph{HST} (i.e., the CCHP or F19 system) and \emph{JWST} TRGB distances yields $H_0$=72.1$\pm 1.1$ km/s/Mpc.  Replacing F19 and H21 distances with those from A22 (where available) raises $H_0$ to 73.3 km/s/Mpc.  The uncertainty in  $H_0$ here and in Table 2 is statistical error only for the purpose of sample comparison with other distance indicators that are calibrated in NGC 4258.  A full error on $H_0$ would include the NGC 4258 distance uncertainty, the error in the SN Hubble diagram intercept,  and any systematic TRGB differences between NGC 4258 and SN hosts as enumerated in A24, Table 3, for a total $\sim$ 1.8 km/s/Mpc.   

Limiting the sample to the same $N=24$ selected by the CCHP \emph{HST}+\emph{JWST} analysis in F25 reduces $H_0$ to 70.7 km/s/Mpc.  Limiting to this CCHP selected sample and further excluding peculiar velocity corrections yields $H_0$=70.2 km/s/Mpc, highly similar to the value of 70.4 km/s/Mpc found by F25 (from the CSP/Snoopy SNe~Ia compilation and also excluding peculiar velocity corrections).  \cite{Peterson_2022ApJ...938..112P} showed that peculiar velocity corrections, derived independently of SNe~Ia or any distance information, significantly reduce the SNe~Ia Hubble diagram dispersion and the overall $\chi^2$ of the fit, making a compelling case for their use.
\cite{Uddin_CSP_2024ApJ...970...72U} found such velocity corrections raise $H_0$ for the CSP SN compilation by 0.55 km/sec/Mpc, so that either our study or the one from F25 yields $\sim$71 with peculiar velocity corrections for this sample of $N=24$.  In this case the difference between the use of Pantheon $+$ and CSP/Snoopy compilation magnitudes produces a difference of $\sim 0.2$ km/s/Mpc.  Similarly, \cite{Uddin_CSP_2024ApJ...970...72U} found little difference in $H_0$ using CSP/Snoopy magnitudes with the \emph{HST} Cepheid distances instead of Pantheon $+$ as either yields $\sim$ 73 km/sec/Mpc.

We investigate why there is a difference in $H_0$ between the CCHP selected sample of 24 and the full sample of 35 SNe~Ia studied here.   The source is seen by considering only the 11 SN~Ia not included in F25 and presented in Table \ref{tab:distances}, which alone yield a higher $H_0$ of 74.1 $\pm 1.7$ km/s/Mpc (statistical only).  The difference between these two independent subsamples of $N=24$ and $N=11$ is 3.4 $\pm$ 2.1 km/s/Mpc (removing common errors), a significance of 1.4 $\sigma$.  The differences in these sub-samples are seen in Figure \ref{fig:H0_histogram}.    

The differences in the combined \emph{HST}+\emph{JWST} samples can be largely traced to those which first appeared in two sub-samples selected for \emph{JWST} follow-up.  As shown in \cite{Riess_JWST_H0_2024arXiv240811770R}, the sub-samples selected for observations with \emph{JWST} produced differences in $H_0$, the same or similar as seen for the same sub-samples using prior \emph{HST} Cepheid measurements. These differences originate from the intrinsic luminosity scatter of the SNe~Ia in the samples, rather than from differences in host distances as measured with either \emph{HST} or \emph{JWST} and with either TRGB or Cepheids. Specifically, we find the CCHP-selected \emph{JWST} sample alone yields 70.2 km/s/Mpc from the TRGB measures. The SH0ES-selected \emph{JWST} sample yields 73.6 km/s/Mpc from the TRGB measures. We reiterate that F25 found a negligible weighted mean difference of $-$0.003 mag (CCHP$-$SH0ES) upon remeasuring a portion of the SH0ES \emph{JWST} sample that is publicly available, so that the difference is not due to TRGB measurement methodology.  Combining both \emph{JWST} samples of 19 SNe~Ia yields 72.4 km/s/Mpc, an apparent reversion to the larger-sample mean.  It is therefore not surprising that the \emph{JWST} sub-sample difference persists in the \emph{HST}+\emph{JWST} sample compiled by F25 of $N=24$ SNe~Ia because this sample excluded the SH0ES \emph{JWST} sample (with no provided reason).  

We also analyze several ``caveat'' samples such as removing all J17 distances (as several studies have been unable to detect the tip in these and F19 did not independently reproduce these measures), using only SNe that pass quality cuts (QC), and removing NGC 1316 (which has the most SNe~Ia, 3, of any single host). We plot the variations in $H_0$ from Table \ref{tab:variants} in a whisker plot in Fig. \ref{fig:H0_whisker}, noting that variants incorporating the full, available SN~Ia sample yields $H_0$ =72.1$-$73.3 $\pm 1.8$ km/s/Mpc. It is expected that increasing sample size naturally leads to a reversion to the mean; conversely, this also means that smaller sample sizes of SN~Ia are susceptible to increased fluctuations in $H_0$.  We caution that when interpreting different $H_0$ values, SN~Ia sample selection and the effects of cosmic variance should be taken into consideration.  Ideally, samples are defined by completeness criteria such as a volume limit, to guard against bias.

\section{Discussion}

\subsection{Our Best Estimate}

We find a best estimate of $H_0$ from TRGB measured in the largest sample of hosts, $N=35$ SNe~Ia, calibrated by NGC 4258, to be 72.1 km/s/Mpc (CCHP TRGB measurement system) to 73.3 km/s/Mpc (EDD TRGB measurement system). These values (which include the Pantheon$+$ SN measurements) are in good agreement with 72.5 km/s/Mpc found from 42 SNe~Ia measured from \emph{HST} Cepheids, as also calibrated by NGC 4258 from R22.  We can reproduce the lower value of $H_0$ found by F25 of 70.4 km/s/Mpc within 0.2 km/s/Mpc by selecting the {\it same} SNe~Ia sub-sample of $N=24$ (i.e. using the same SNe~Ia selection as in F25, lowering $H_0$ by 1.4 km/sec/Mpc) and by excluding peculiar velocity corrections used in the Pantheon $+$ sample (lowering $H_0$ by 0.6 km/s/Mpc) \citep{Peterson_2022ApJ...938..112P} which results in 70.2 km/s/Mpc.
The difference between Pantheon$+$ SN magnitudes and the CSP/Snoopy compilation favored by F25 appear to produce a difference at the 0.3 km/s/Mpc level.

The largest difference between F25 and here comes from 11 SNe~Ia not included by F25, most from the \emph{JWST} SH0ES sample (whose TRGB distance measures were confirmed by F25 for a subset of galaxies to 0.003 mag in the mean).  Just using the $N=24$ sample from F25 and the \emph{JWST} SH0ES sample as measured by F25 would raise $H_0$ to 71.8 km/sec/Mpc.
The above SNe~Ia sub-sample differences in $H_0$ may be attributed to the statistics of small-samples; a $\sim$ 2$\sigma$ difference is not very unusual (F25 does not provide the 
method for selecting \emph{JWST} TRGB host targets).
However, we see no rational for not including all 35 TRGB calibrations of SNe~Ia in a best estimate of $H_0$ since they are consistently obtained and measured.   Specifically, the \emph{JWST} SH0ES sample was measured in F25 yet still excluded.  

As sample sizes increase, we see no relief coming into focus for the persistent Hubble tension.  F25 suggests an unorthodox avenue for reducing the significance of the Hubble tension; increase the uncertainty in the SH0ES measurement of $H_0$.  Specifically, F25 proposes that the uncertainty in $H_0$ could be increased by adding to the standard error propagation the size of historical changes (i.e., available improvements) to measured quantities between some past studies.\footnote{For example F25 adds to the uncertainty 
in R22 the size of the mean change, 0.03 mag, between SN magnitudes from \cite{Scolnic_2015ApJ...815..117S} and \cite{Scolnic_Pantheon_2022ApJ...938..113S}.  The origin of this shift, as explained in R22, is the availability of additional SN surveys, doubling the mean number of light curves per SN calibrator.  F25 also shifts the value of $H_0$ in relation to measured differences between the ground calibration of LMC Cepheids available in \cite{Riess_2016ApJ...826...56R} and their \emph{HST}-calibration in \cite{Riess_2019ApJ...876...85R}}.  We have not seen such changes propagated as errors for new iterations of other experiments (such as for the CMB WMAP, Planck, ACT or SPT series) and do not think it is sensible to ``inherit'' uncertainty from past iterations of an experiment.   Improvement through iteration is generally found to reduce uncertainty, not accumulate it. Fortunately, the impact of SN sub-sample differences on $H_0$ should continue to diminish as the sample size of TRGB-SNe~Ia calibrators continues to expand, provided all available statistics are employed. 

\subsection{Negligible Supernova Compilation Differences}

As discussed above, we can reproduce the $H_0=70.4$ km/s/Mpc determined by F25 to within 0.2 km/s/Mpc (here $H_0=70.2$ km/s/Mpc) by limiting the SN sample to the same selections ($\Delta H_0$ of 1.1 km/s/Mpc) and neglecting corrections for peculiar velocities ($\Delta H_0$ of 0.6 km/s/Mpc).    F25 omits the peculiar velocity correction, used in the CSP study of \cite{Uddin_CSP_2024ApJ...970...72U} and instead places this amount in an additional uncertainty (denoted as $\sigma_{SN}$) of 1 km/s/Mpc.  The difference of 0.2 km/s/Mpc between F25 and here is attributed to the only remaining difference, the SN compilations used, CSP/Snoopy (F25) vs Pantheon$+$ (here).

Still, H25 argues that issues with the Pantheon$+$ compilation from \cite{Scolnic_2023ApJ...954L..31S} is a source of differences in $H_0$ measured between different analyses. To reach this conclusion, H25 compares SN $M_B$ between varying membership of the CCHP \emph{JWST} subsample and of the \emph{HST}+\emph{JWST} sample, and finds differences ranging from 0.03-0.06 mag, depending on the SN compilation and sample membership used.  H25 finds the largest and most significant difference for Pantheon$+$ and ascribes this to an issue with Pantheon$+$.  However, in Table \ref{tab:variants} we find a difference between the ``Only CCHP JWST" sample of $N=11$ and the ``CCHP Selected" \emph{HST}+\emph{JWST} $N=24$ sample from Pantheon$+$ is $\Delta M_B=0.029$~mag, nearly identical to the CSP(I+II) results found by H25 of 0.027 mag for the {\it same} objects.  (H25 limited the Pantheon$+$ analysis to $N=17$ objects and does not indicate which ones.  Our finding using the same $N=24$ sample indicates the difference between the $N=17$ and $N=24$ samples raised the difference to 0.06 mag rather than a consequence of changing SN compilations).  So from this same comparison, we see no significant difference between the SN compilations.  This is consistent with the negligible differences in $H_0$ found by \cite{Uddin_CSP_2024ApJ...970...72U} when using the same host distances as R22 and the CSP/Snoopy compilation. We also note that this sample comparison should not presume the subsample difference is zero (and that any significant finding of a difference is the fault of the measurement), because by every measure employed these subsamples appear systematically unequal.  

We also note that there are good reasons to favor a SN compilation composed of many SN surveys for the distance ladder due to the consistent assembly of SN on both rungs \citep{Brownsberger_2023ApJ...944..188B}.    For example, \cite{Uddin_CSP_2024ApJ...970...72U}, F25 and H25 include only CSP SNe for their Hubble flow sample, but the majority of SNe in the calibrator rung are not from the CSP survey.  Also, the Pantheon+ analysis homogeneously recalibrates all surveys by tying reference stars to a common photometric system (such as Pan-STARRS) to avoid inconsistency in calibration, whereas these steps were not applied in the ``CSP" sample.  It is also noteworthy that Pantheon$+$ contains the CSP survey and benefits from the averaging of up to 4 sources of light curves for SN calibrators.   This becomes important for the consideration of possible outlier SNe.  For example, H25 argues one calibrator, SN 2007af, in R22 is an outlier (formally it is 2.97 $\sigma$ off and its retained because R22 used a 3.3$\sigma$ automated threshold based on Chauvenet's criterion and the number of data in that study) and excluding it can shift $H_0$ by $\sim$ 0.5 km/s/Mpc. Because Pantheon$+$ has 4 independent light curves for this SN (CfA3, LOSS, CSP and Swift) we can be confident the SN magnitude is reliable.  As for it's host distance, in Table 3 there are multiple distance measures for its host, NGC 5584, from TRGB of 31.82 $\pm 0.10$ mag (\emph{HST}; JL17), 31.80 $\pm 0.11$ mag (\emph{JWST}; L24), 31.85 $\pm 0.05$ mag (\emph{JWST}; F25), and for Cepheids 31.76 $\pm 0.06$ mag (\emph{HST}; R22) and 31.84 $\pm 0.03$ mag (\emph{JWST}; R24), so we can be confident about the distance measurement.  Therefore its difference from the mean would be primarily intrinsic to the SN and the population.  Whether the SN is an intrinsic outlier then depends on the size of the sample in which it is found; if the sample is all SN used on the distance ladder as the automated clipping criterion assumes, several hundred SNe, it is probably not.

It is useful then to consistently compare the Pantheon$+$ and CSP sample in terms of inferred $H_0$. Using Cepheid calibrators from R22, R22 find $H_0 = 73.0 \pm 1.0$ km/s/Mpc using the Pantheon$+$ sample for 42 SNe~Ia.  \cite{Uddin_CSP_2024ApJ...970...72U} find for 25 SNe Ia in the CSP/Snoopy compilation a value of $H_0 = 72.55 \pm 0.76$ km/s/Mpc when calibrating the SN with B band magnitudes, and $H_0 = 73.22 \pm 0.75$ km/s/Mpc when calibrating with H band magnitudes. We also note a third analysis from \cite{Dhawan_BayeSN_H0_2023MNRAS.524..235D} that also measures $H_0$ with the full sample from R22 and F19,  where they find $H_0 = 74.82 \pm 1.0$ km/s/Mpc. For TRGB calibrators from F19 with 19 supernovae, we measure here $H_0 = 70.9 \pm 1.5$ km/s/Mpc using Pantheon+, \cite{Uddin_CSP_2024ApJ...970...72U} find $H_0 = 70.32 \pm 0.68$ km/s/Mpc and $70.99 \pm 0.85$ km/s/Mpc from B and H respectively and without peculiar velocity corrections (reducing $H_0$ by $\sim$ 0.6 km/s/Mpc), and \cite{Dhawan_BayeSN_H0_2023MNRAS.524..235D} finds $70.92 \pm 1.14$ with peculiar velocity corrections. The uncertainties given here are statistical only; \cite{Dhawan_BayeSN_H0_2023MNRAS.524..235D} estimates an additional systematic uncertainty of 0.84 km/s/Mpc for the Cepheid case and 1.49 km/s/Mpc for the TRGB case.  We therefore find great consistency between these three analyses, with the only slightly discrepant value from \cite{Dhawan_BayeSN_H0_2023MNRAS.524..235D} when including the full set of 42 Cepheid– SN Ia calibrators, as they find a higher value of $H_0=74.82 \pm 1.0$ km/s/Mpc. Simply put, there is no indication that Pantheon$+$ is pulling $H_0$ relative to these other SN compilations when the source of the SN compilation is the only substitution made.

We note a few areas where further work is needed to enhance our understanding of distance ladder data.   H25 and F25 used the same sample from \cite{Uddin_CSP_2024ApJ...970...72U} with the same TRGB distances but produce a difference of $\sim 1$ km/s/Mpc.  H25 explains this may be due to the different Snoopy standardization parameters for SNe~Ia used between the analyses.  F25 refits these parameters given the new TRGB distances, whereas H25 uses specific values from \cite{Uddin_CSP_2024ApJ...970...72U} trained on a smaller, prior TRGB subset. The sensitivity of the standardization parameters to the calibration distances is somewhat surprising as only a small sample of data from  the second calibrator rung changed and not the larger Hubble flow rung.   One place where $H_0$ sensitivity should arise is from the correction for the `mass-step', the difference in standardized brightness for SNe in different types of host galaxies.  For both the values that F25 and H25 use, the split point is at a higher mass than what is typically used \cite{2010MNRAS.406..782S} and a slope rather than a step is used.  This may underestimate the step size and affect the other standardization parameters, increasing sensitivity to the TRGB distances.

Finally, F25 claims a 3$\sigma$ trend is evident in a plot (Figure B1) of the R22 Cepheid distances vs R22 $M_B$ suggesting a correlation between SN mag and distance.  This same data was compared in R22 where the trend was found there to be 1.5 $\sigma$ and hence not significant.  Close study of the new figure B1 in F25 shows that the plotted errors are not the same as those in R22 (Table 6) appearing far smaller in the F25 plot (e.g., for 2009Y at $\mu=33.1$ and $M_B=-19.6$, where R22 lists an uncertainty in distance of 0.2 mag and in $M_B$ of 0.24 mag while the F25 plot has an error in $M_B$ of 0.1 mag and in $\mu$ of zero), with several such points at large distance appearing to cause the difference.  We would like to understand if the R22 data was transformed in some way and if this causes the difference in claimed trend.  A larger study of the linearity of the HST Cepheid distances using \emph{JWST} and multiple indicators in \cite{Riess_JWST_H0_2024arXiv240811770R} shows an even less significant trend of $<$ 1 $\sigma$.

\begin{sidewaystable}[htbp]
\centering
\tiny
\caption{Host Galaxy Distances, SN~Ia magnitudes, and $H_0$ values}
\label{tab:distances}
\begin{tabular}{cccccccccccccccccccccc}
\hline
\textbf{SN \#} & \textbf{Galaxy} & $\mu^{\mathrm{TRGB}}_{\mathrm{HST}}$ & $\sigma$* & Src & $\mu^{\mathrm{TRGB}}_{\mathrm{HST}}$(A22) & $\sigma$* & $\mu^{\mathrm{TRGB}}_{\mathrm{JWST}}$ & $\sigma$* & Src & $\mu^{\mathrm{TRGB}}_{\mathrm{JWST}}$(F24) & $\sigma$* & $\bar{\mu}$ & $\sigma$* & SNe~Ia & P+SN [mag] & $\sigma$ & $M_B$ & $\sigma$ & $H_0$ & $\mu^{\mathrm{Ceph}}_{\mathrm{HST}}$ & $\sigma$ \\
\hline
1 & M101 & 29.08 & 0.04 & F19 & 29.08 & 0.05 & 29.151 & 0.042 & F24 & -- & -- & 29.11&0.03& 2011fe & 9.78 & 0.12 & -19.33 & 0.12 & 71.1 & 29.188 & 0.055 \\
2 & M66 & 30.22 & 0.04 & F19 & 30.22 & 0.05 & -- & -- & -- & -- & -- & 30.22&0.04&1989B & 10.98 & 0.15 & -19.24 & 0.16 & 74.2 & -- & -- \\
3 & M96 & 30.31 & 0.04 & F19 & 30.20 & 0.05 & -- & -- & -- & -- & -- & 30.31&0.04&1998bu & 11.00 & 0.15 & -19.31 & 0.16 & 71.8 & -- & -- \\
4 & N1309 & 32.50 & 0.07 & J17 & -- & -- & -- & -- & -- & -- & -- & 32.50&0.07&2002fk & 13.20 & 0.12 & -19.3 & 0.14 & 72.2 & 32.552 & 0.069 \\
5 & N1316 & 31.46 & 0.04 & F19 & 31.36 & 0.07 & -- & -- & -- & -- & -- & 31.46&0.04&1980N & 12.002 & 0.097 & -19.46 & 0.10 & 67.1 & -- & -- \\
6 & N1316 & 31.46 & 0.04 & F19 & 31.36 & 0.07 & -- & -- & -- & -- & -- & 31.46&0.04&2006dd & 11.94 & 0.108 & -19.52 & 0.12 & 65.2 & -- & -- \\
7 & N1316 & 31.46 & 0.04 & F19 & 31.36 & 0.07 & -- & -- & -- & -- & -- & 31.46&0.04&1981D & 11.61 & 0.23 & -19.85 & 0.23 & 56.0 & -- & -- \\
8 & N1365 & 31.36 & 0.05 & F19 & 31.41 & 0.05 & 31.366 & 0.069 & F24 & -- & -- & 31.36&0.04&2012fr & 11.90 & 0.09 & -19.46 & 0.1 & 67.0 & 31.378 & 0.061 \\
9 & N1380 & -- & -- & -- & -- & -- & 31.397 & 0.072 & A24 & -- & -- & 31.397&0.07&1992A & 12.095 & 0.135 & -19.30 & 0.15 & 72.1 & -- & -- \\
10 & N1404 & 31.36 & 0.06 & H21 & 31.29 & 0.07 & -- & -- & -- & -- & -- & 31.36&0.06&  2011iv & 11.974 & 0.099 & -19.39 & 0.12 & 69.4 & -- & -- \\
11 & N1404 & 31.36 & 0.06 & H21 & 31.29 & 0.07 & -- & -- & -- & -- & -- &31.36&0.06& 2007on & 12.46 & 0.19 & -18.9 & 0.20 & 86.8 & -- & -- \\
12 & N1448 & 31.32 & 0.06 & F19 & 31.33 & 0.06 & 31.39 & 0.07 & L24 & 31.321 & 0.049 &31.35&0.05& 2001el & 12.254 & 0.136 & -19.1 & 0.14 & 79.3 & 31.298 & 0.051 \\
13 & N1448 & 31.32 & 0.06 & F19 & 31.33 & 0.06 & 31.39 & 0.07 & L24 & 31.321 & 0.049 &31.35&0.05& 2021pit & 11.752 & 0.20 & -19.6 & 0.21 & 62.9 & -- & -- \\
14 & N1559 & -- & -- & -- & -- & -- & 31.51 & 0.05 & L24 & 31.491 & 0.051 & 31.51&0.05&2005df & 12.141 & 0.086 & -19.37 & 0.1 & 69.9 & 31.5 & 0.071 \\
15 & N2442 & -- & -- & -- & -- & -- & 31.646 & 0.097 & F24 & -- & -- &31.65&0.10&2015F & 12.23 & 0.09 & -19.42 & 0.13 & 68.4 & 31.459 & 0.073 \\
16 & N2525 & -- & -- & -- & -- & -- & 31.81 & 0.09 & L24 & -- & -- & 31.81&0.09 &2018gv & 12.728 & 0.074 & -19.08 & 0.12 & 79.8 & 32.059 & 0.105 \\
17 & N3021 & 32.22 & 0.05 & J17 & -- & -- & -- & -- & -- & -- & -- & 32.22&0.05&1995al & 12.97 & 0.12 & -19.25 & 0.13 & 73.8 & 32.473 & 0.162 \\
18 & N3370 & 32.27 & 0.05 & J17 & -- & -- & 32.19 & 0.08 & L24 & -- & -- & 32.25&0.04&1994ae & 12.937 & 0.082 & -19.31 & 0.09 & 71.8 & 32.132 & 0.062 \\
19 & N3447 & -- & -- & -- & -- & -- & 31.92 & 0.09 & L24 & -- & -- & 31.92&0.09&2012ht & 12.736 & 0.089 & -19.18 & 0.13 & 76.1 & 31.947 & 0.049 \\
20 & N3972 & -- & -- & -- & -- & -- & 31.747 & 0.068 & F24 & -- & -- & 31.75&0.07&2011by & 12.55 & 0.09 & -19.20 & 0.11 & 75.7 & 31.644 & 0.096 \\
21 & N3982 & 31.50 & 0.13 & Here & -- & -- & -- & -- & -- & -- & -- & 31.50&0.13&1998aq & 12.252 & 0.078 & -19.25 & 0.15 & 73.9 & 31.736 & 0.08 \\
22 & N4038 & 31.68 & 0.05 & J17 & 31.68 & 0.14 & 31.645 & 0.078 & F24 & -- & -- & 31.67&0.04&2007sr & 12.41 & 0.11 & -19.26 & 0.12 & 73.5 & 31.612 & 0.121 \\
23 & N4414 & 31.24 & 0.09 & Here & -- & -- & -- & -- & -- & -- & -- & 31.24&0.09&2021J & 12.1046 & 0.14 & -19.14 & 0.17 & 77.8 & -- & -- \\
24 & N4424 & 31.00 & 0.06 & F19 & 31.00 & 0.07 & 30.926 & 0.03 & F24 & -- & -- &30.94&0.03& 2012cg & 11.487 & 0.19 & -19.45 & 0.19 & 67.2 & 30.854 & 0.133 \\
25 & N4457 & 31.05 & 0.10 & Here & -- & -- & -- & -- & -- & -- & -- & 31.05&0.1&2020nvb & 11.85 & 0.14 & -19.20 & 0.17 & 75.6 & -- & -- \\
26 & N4526 & 31.00 & 0.07 & F19 & 30.99 & 0.05 & -- & -- & -- & -- & -- & 31.00&0.07&1994D & 11.532 & 0.093 & -19.47 & 0.12 & 66.8 & -- & -- \\
27 & N4536 & 30.96 & 0.05 & F19 & 31.01 & 0.13 & 30.923 & 0.052 & F24 & -- & -- & 30.94&0.04&1981B & 11.55 & 0.13 & -19.39 & 0.13 & 69.2 & 30.87 & 0.061 \\
28 & N4639 & 31.79 & 0.09 & Here & -- & -- & 31.774 & 0.073 & F24 & -- & -- & 31.78&0.06&1990N & 12.45 & 0.12 & -19.33 & 0.13 & 71.2 & 31.823 & 0.091 \\
29 & N4666 & 30.90 & 0.05 & Here & -- & -- & -- & -- & -- & -- & -- & 30.90&0.05&ASASSN-14lp & 11.585 & 0.13 & -19.32 & 0.14 & 71.7 & -- & -- \\
30 & N5584 & 31.82 & 0.10 & J17 & -- & -- & 31.8 & 0.11 & L24 & 31.851 & 0.053 &31.81&0.07& 2007af & 12.804 & 0.079 & -19.01 & 0.11 & 82.6 & 31.766 & 0.062 \\
31 & N5643 & 30.48 & 0.08 & H21 & 30.42 & 0.07 & 30.58 & 0.06 & L24 & 30.599 & 0.057 & 30.54&0.05&2013aa & 11.25 & 0.08 & -19.29 & 0.09 & 72.4 & 30.553 & 0.063 \\
32 & N5643 & 30.48 & 0.08 & H21 & 30.42 & 0.07 & 30.58 & 0.06 & L24 & 30.599 & 0.057 & 30.54&0.05&2017cbv & 11.21 & 0.08 & -19.33 & 0.09 & 71.0 & -- & -- \\
33 & N5861 & -- & -- & -- & -- & -- & 32.10 & 0.11 & L24 & -- & -- & 32.1&0.11&2017erp & 12.945 & 0.107 & -19.16 & 0.15 & 77.2 & 32.232 & 0.105 \\
34 & N7250 & -- & -- & -- & -- & -- & 31.629 & 0.047 & F24 & -- & -- & 31.63&0.05 &2013dy & 12.28 & 0.18 & -19.35 & 0.19 & 70.6 & 31.642 & 0.13 \\
35 & N7814 & 30.86 & 0.07 & D21 & -- & -- & -- & -- & -- & -- & -- & 30.86&0.07&2021rhu & 11.92 & 0.15 & -18.94 & 0.17 & 85.2 & -- & -- \\
\hline
\end{tabular}
\begin{tablenotes}
\footnotesize
\item \textbf{Notes: Compilation of TRGB distances to, Cepheid distances to, and SN~Ia \& $H_0$ for, the host galaxies used in this study. Columns from left to right are: SN~Ia number count, host galaxy names, \emph{HST} TRGB distances, uncertainties on those distances, references for those distances, \emph{HST} TRGB distances from A22, uncertainties on those distances, \emph{JWST} TRGB distance, uncertainties on those distances, sources for those distances, \emph{JWST} TRGB distances from \cite{Freedman_JWST_H0_2024arXiv240806153F}, which use underlying images from \emph{JWST} Cycle 1 program GO-1685 (CCHP reanalysis of SH0ES data in F25), uncertainties on those distances, SN~Ia name, SN~Ia magnitudes from Pantheon+ \cite{Scolnic_Pantheon_2022ApJ...938..113S}, uncertainties on those SN~Ia magnitudes, SN~Ia $M_B$ using the weighted mean \emph{HST} and \emph{JWST} TRGB distances (where applicable), uncertainties on those $M_B$, $H_0$ using that single SN~Ia, \emph{HST} Cepheid distance (anchored to N4258 only) from \cite{Riess_JWST_H0_2024arXiv240811770R}, and uncertainties on those distances.  If more than one source exists in P$+$ for a SN mag, we use the mean (from Table 6 of R22 but different than \citep{Scolnic_2023ApJ...954L..31S} who used the IDL median.  
For the J17 entries, we use the N4258 calibrated distances, with the zero-point adjusted to match that as given in F19. *N5643 has additional measurements from \cite{Freedman_JWST_H0_2024arXiv240806153F}, for instance, of $M^{F115W}_{TRGB}=30.643 \pm 0.071$; we use the one from L24 due to the smaller color dependence \citep{Newmann_JWST_2024arXiv240603532N} in the \emph{F090W} band compared to the \emph{F115W} band used in \cite{Freedman_JWST_H0_2024arXiv240806153F} . *The distance to N1380 from \cite{Anand_Fornax_2024ApJ...973...83A} has its error updated in \cite{Jensen_2025arXiv250215935J}, which we list here. The distances to NGC 1380 and NGC 7814 both include the NGC 4258 TRGB measurement error and maser distance error. Distance uncertainties from J17, F19, H21, and F25 are presumed not to include any component of error from NGC 4258 distance or tip measure. TRGB distances labeled `Here' and from L24 include the TRGB measurement error in NGC 4258.  }
\end{tablenotes}
\end{sidewaystable}

\begin{acknowledgments}

We are indebted to all of those who spent years and even decades bringing {\it JWST} to fruition. We thank Stefano Casertano, Wenlong Yuan, and Louise Breuval for helpful conversations. SL is supported by the National Science Foundation Graduate Research Fellowship Program under grant number DGE2139757. GSA acknowledges financial support from {\it JWST} GO-1685 and JWST GO-2875. DS is supported by Department of Energy grant DE-SC0010007, the David and Lucile Packard Foundation, the Templeton Foundation and Sloan Foundation. 

This research is based in part on observations made with the NASA/ESA Hubble Space Telescope obtained from the Space Telescope Science Institute, which is operated by the Association of Universities for Research in Astronomy, Inc., under NASA contract NAS 5–26555, and the NASA/ESA/CSA James Webb Space Telescope.  The data were obtained from the Mikulski Archive for Space Telescopes at the Space Telescope Science Institute, which is operated by the Association of Universities for Research in Astronomy, Inc., under NASA contract NAS 5-03127 for \emph{JWST}. These observations are associated with programs $\#$17079  and $\#$16453. This research has made use of the NASA/IPAC Extragalactic Database (NED),
which is operated by the Jet Propulsion Laboratory, California Institute of Technology,
under contract with the National Aeronautics and Space Administration.

\end{acknowledgments}

\facilities{HST(ACS), JWST(NIRCam)} 
\software{astropy \citep{Astropy_2013A&A...558A..33A},  
   DOLPHOT
   \citep{Dolphin_2000PASP..112.1383D, Dolphin_2002MNRAS.332...91D, Dolphin_2016ascl.soft08013D}
          }

\eject

\bibliography{main}{}
\bibliographystyle{aasjournalv7}

\end{document}